\documentclass[twocolumn]{aa}
\usepackage{graphicx}
\usepackage{txfonts,times}
\begin{document}
   \title{Tiny \ion{H}{i} Clouds in the Local ISM}

   \author{Robert Braun
          \inst{1}
          \and
          Nissim Kanekar\inst{2}
          }

   \offprints{R. Braun}

   \institute{ASTRON, Postbox 2, 7990 AA Dwingeloo, The Netherlands\\
              \email{braun@astron.nl}
         \and
             National Radio Astronomy Observatory, Socorro, NM 87801, USA\\
             \email{nkanekar@aoc.nrao.edu}
             }

   \date{Received ; accepted }

\abstract{We report deep, high spectral resolution WSRT \ion{H}{i} 21cm
observations of four high latitude compact radio sources, that have revealed a
new population of tiny, discrete clouds in the diffuse ISM, with peak optical
depths $\tau \sim 0.1 - 2$\%, \ion{H}{i} column densities of 0.4--8 $\times
10^{18}$~cm$^{-2}$ and core temperatures of 20--80~K.  Imaging
detections confirm these low column densities and imply linear core
dimensions of a few thousand~AU, assuming a distance of 100~pc. The physical
origin of these tiny \ion{H}{i} structures and their distribution in the ISM
is at present unknown. Further observations will be required to determine
whether they are a ubiquitous component of the ISM.

   }

   \maketitle
%

\section{Introduction}

Despite more than three decades of \ion{H}{i} 21cm absorption \& emission
studies, we still do not have a good understanding of the physical
characteristics of the atomic component, one of the most important phases of
the interstellar medium (ISM). Crucial observational lacunae include the
morphology of neutral hydrogen ``clouds'', the time-scales of their formation
and destruction (i.e. whether they are transient or relatively stable
structures), the \ion{H}{i} column density distribution function, the nature
of the equilibrium between ``cold'' and ``warm'' phases, etc.  In recent
times, there have been several indications of a surprising degree of
small-scale structure in the atomic ISM. Perhaps the first of these was the
observation of spatially variable \ion{H}{i} absorption towards compact radio
sources (e.g. Dieter et al.~\cite{diet76}; Davis et al.~\cite{davi96}; Faison
et al.~\cite{fais98}). In the most extreme cases, for example towards 3C138,
there is evidence for changes of as much as $\Delta\tau$~=~0.1 in the
\ion{H}{i} opacity on transverse scales of 20~AU (Faison et
al.~\cite{fais98}). The simplest interpretation of these observations requires
large variations in the volume density,
$\Delta$n$_{\ion{H}{i}}~\sim~10^5$~cm$^{-3}$, assuming that all of the other
relevant variables (specifically, the pathlength and the temperature) are kept
fixed.  However, as argued by Deshpande~(\cite{desh00}), realistic ISM
structure functions can lead to large variations of $\tau$ with small angular
offsets, simply from statistical fluctuations in the effective pathlength with
position. Further evidence for small-scale atomic structure stemmed from
searches for time variability in the \ion{H}{i} absorption seen toward pulsars
(e.g. Frail et al.~\cite{frai94}). However, the early claims for ubiquitous
and significant time variations in $\tau$ have not been confirmed in more
recent studies. (e.g. Johnston et al.~\cite{john03}; Stanimirovic et
al.~\cite{stan03}).  A third line of evidence has come from observations of
\ion{Na}{I} observation toward nearby pairs of stars (eg. Watson \&
Meyer~\cite{wats96}; Points et al.~\cite{poin04} and references therein),
suggesting discrete absorption features in the local ISM with highly variable
properties on scales of hundreds of AU. However, Points et al.~(\cite{poin04})
note that \ion{Na}{I} is not the dominant ion in diffuse clouds and it is thus
possible that the above structures are not distinct physical entities but
merely reflect fluctuations in local physical conditions, such as the
temperature, pressure, electron density, etc.

\section{Observations}
\label{sec:obs}

We have recently undertaken a series of extremely sensitive \ion{H}{i}
absorption observations towards bright compact extra-galactic radio sources
near the North Galactic Pole (NGP), with the Westerbork Synthesis Radio
Telescope (WSRT). The initial motivation for these observations was the
detection of \ion{H}{i} absorption from a Warm Neutral Medium (WNM), even for
temperatures as high as 10$^4$~K.  Such an experiment was prompted by the
detection of wide \ion{H}{i} absorption by Kanekar et al.~(\cite{kane03}),
with velocity spreads corresponding to an equivalent temperature of 3500~K;
these observations reached a $1\sigma$ optical depth sensitivity of $\sim
6\times 10^{-4}$ per 1~km/s velocity channel. The NGP region was chosen to
ensure that the lines-of-sight through the Galactic disk might be as short as
possible and, therefore, relatively simple. Sensitive, high velocity
resolution absorption spectra were acquired towards 3C286 (14.7~Jy, 12 hr
integration), 3C287 (7~Jy, 9 hr), 4C+32.44 (5~Jy, 11 hr), and B2~1325+32
(1.4~Jy, 11~hr), by observing in an in-band frequency-switching mode,
utilising a 1~MHz throw every 5 minutes inside a total bandwidth of 2.5~MHz,
with a channel width of 0.5~km/s.  The in-band frequency switching allowed
exceptionally good band-pass calibration while providing 100\% of observing
time on-source. We achieved {\sc RMS} optical depth sensitivities $\Delta
\tau$ of 1.8$\times 10^{-4}$ (for 3C286), 2.9$\times 10^{-4}$ (for 3C287) and
4.0$\times 10^{-4}$ (for 4C+32.44) at 1~km/s velocity resolution.  These are
the most sensitive \ion{H}{i} 21cm absorption measurements of which we are
aware. A somewhat lower sensitivity ($\Delta \tau \sim 0.0028$ per 1~km/s) was
obtained towards B2~1325+32, due to its lower flux density. Finally, we
also used the simultaneously acquired total power spectra of the WSRT single
dishes to derive sensitive \ion{H}{i} emission profiles (with effective
integration times of more than 100 hr) towards all four sources. The flux
scale is based on the SEFD of the WSRT dishes at 1400 MHz of 300$\pm$10
Jy/Beam. We emphasise that only a constant term has been subtracted from each
spectrum (both absorption and emission) and not even a first order polynomial,
let alone a higher order spectral baseline.

\begin{figure}[t!]
\centering
\includegraphics[width=8.cm]{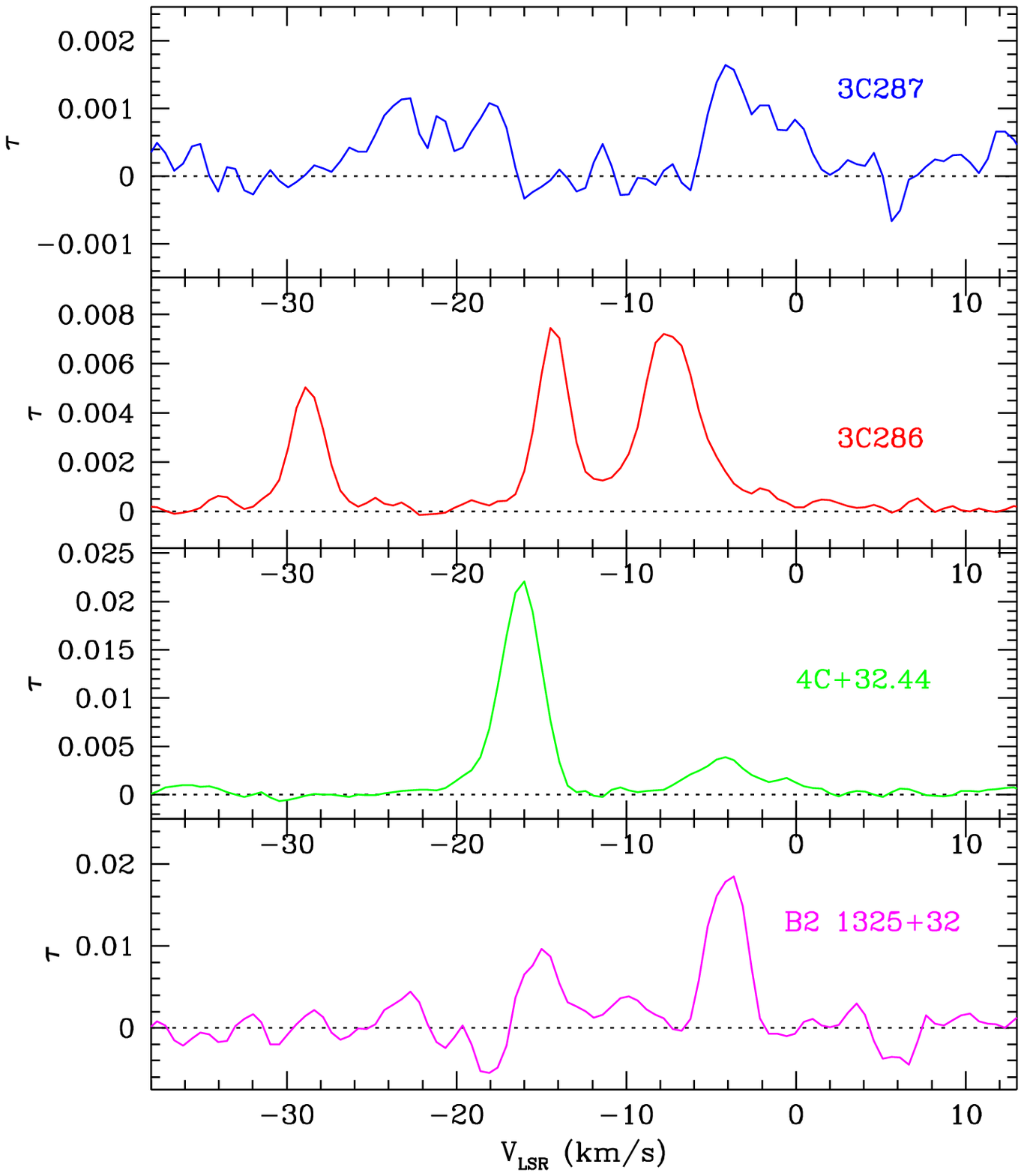}\\
\includegraphics[width=8.cm]{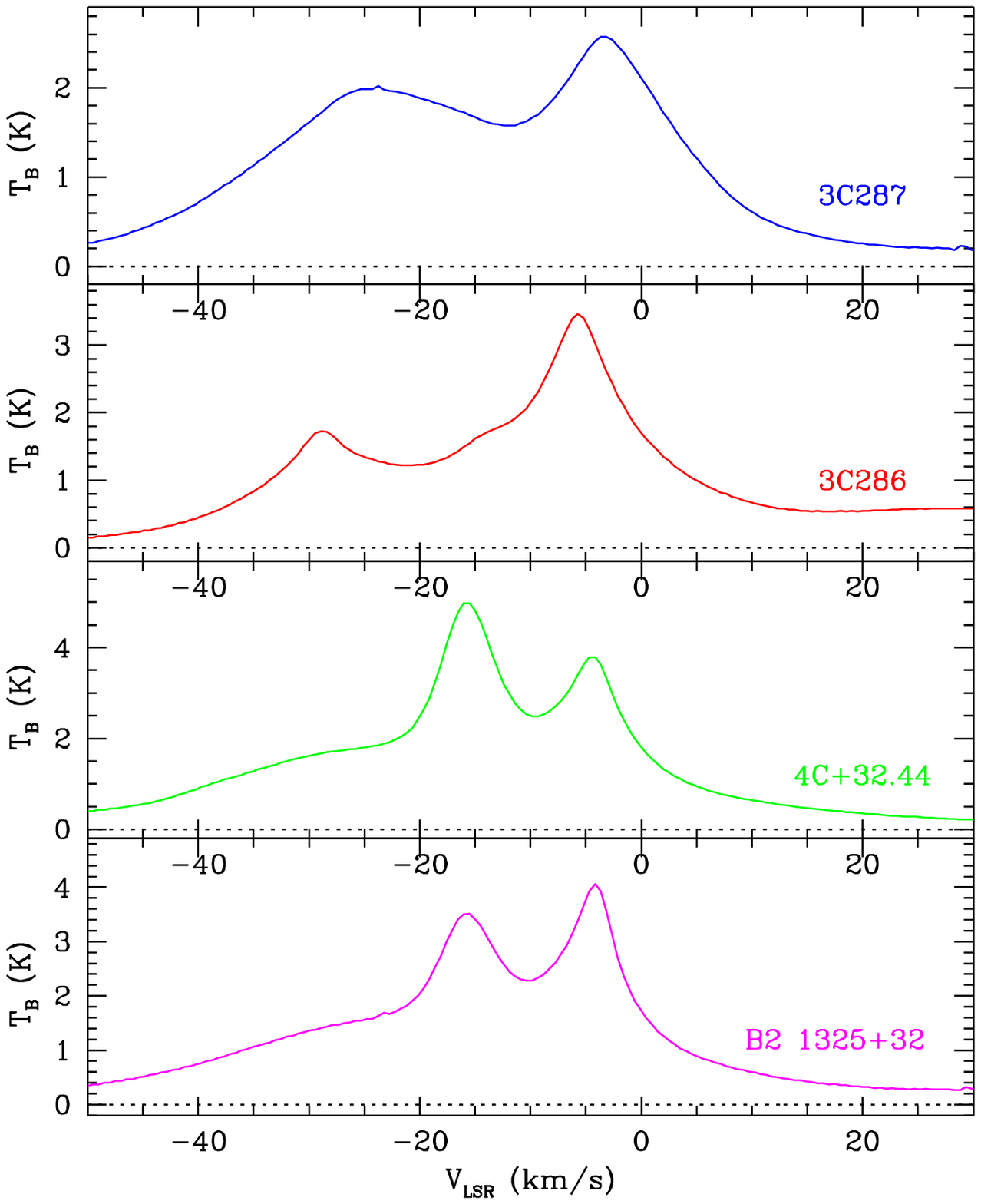}
\caption{\ion{H}{i} absorption (top) and total power emission (bottom)
  spectra after Hanning smoothing to 1~km/s velocity resolution.} 
\label{fig:allspec}
\vskip-0.2cm
\end{figure}

\section{Results}
\label{sec:results}

Since the primary motivation for the experiment was a search for broad,
shallow absorption along ``simple'' high latitude lines of sight,
we were surprised to find instead that multiple, narrow (FWHM $\sim$ 2--3~km/s)
absorption features were detected at discrete line-of-sight velocities, with
peak opacities in the range 0.1--2\%. A representative cool phase spin
temperature of $100$~K (corresponding to a thermal FWHM of 2.2~km/s) yields
extremely low \ion{H}{i} column densities, 0.4--8 $\times 10^{18}$~cm$^{-2}$,
for these components. The detection of discrete absorption features was
particularly surprising since the peak emission brightness seen in these
directions (with the 35\arcmin total power beam) is only between 2 and 5~K.
The absorption and emission spectra towards the four sources are compared in
the top and bottom panels of Figure~\ref{fig:allspec}. It is
striking that moderately nearby lines-of-sight (separation $< 105'$ between
3C286 and 4C+32.44) show essentially uncorrelated absorption spectra, and only
weakly correlated emission features. Only in the case of 4C+32.44 and
B2~1325+32, which have an angular separation of only $15'$, is there a hint of
similar velocity components having been detected. In all cases, the emission
spectra show wide wings extending out to $\pm 50$~km/s, while the absorption
is restricted to the velocity range $0$ to $-30 $~km/s. The very broadest
emission in the spectra of Fig.~\ref{fig:allspec}(b) is actually due to
so-called stray radiation, \ion{H}{i} emission attenuated by 30 to 40 dB, but
still detected in the far side-lobes of the telescope response when the bright
Galactic plane happens to be above the local horizon.

\begin{figure}
\centering
\includegraphics[width=4.cm]{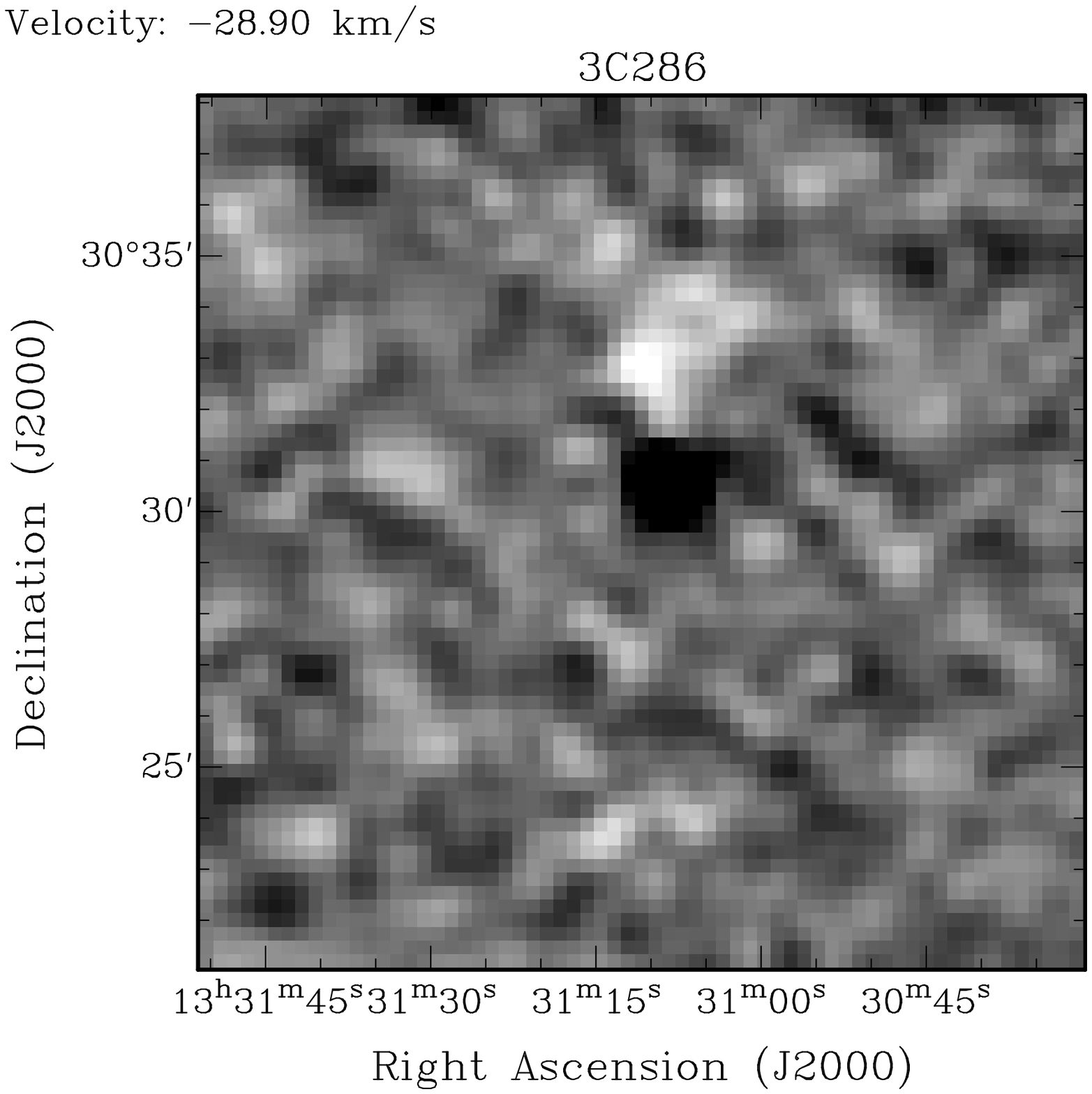} 
\hskip0.3cm
\includegraphics[width=3.8cm]{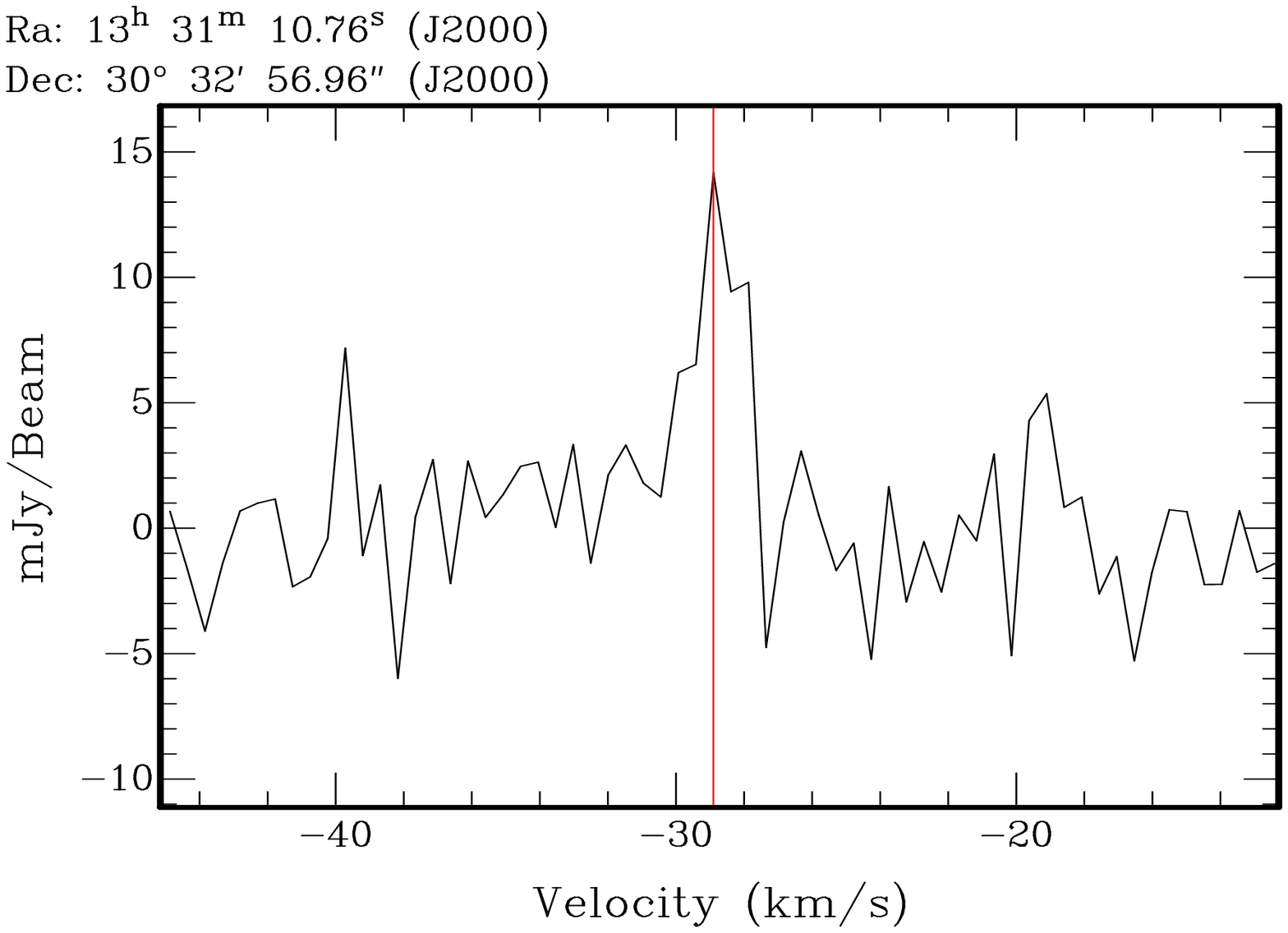}\\
\vskip0.1cm
\includegraphics[width=4.cm]{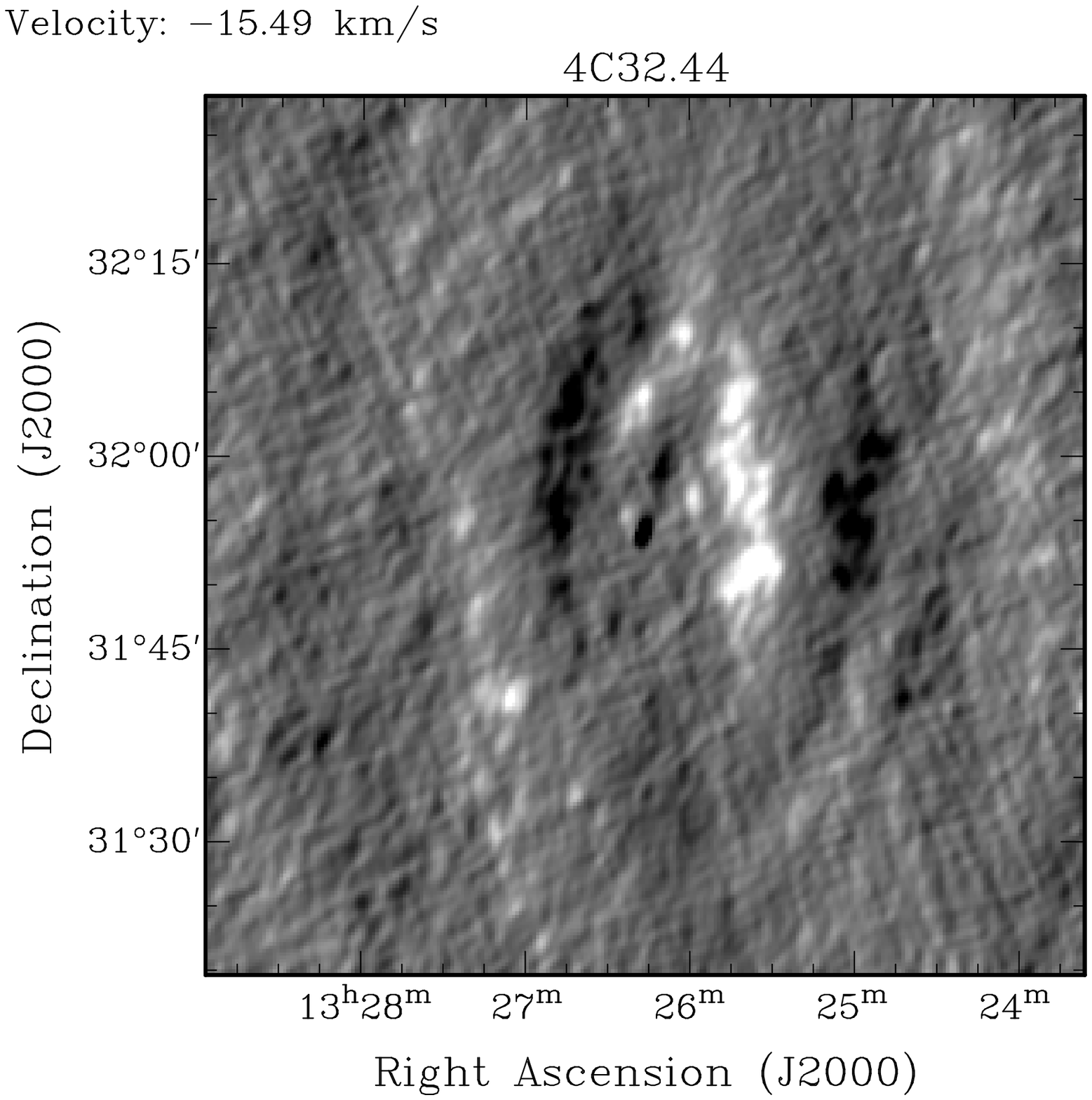}
\hskip0.3cm
\includegraphics[width=3.8cm]{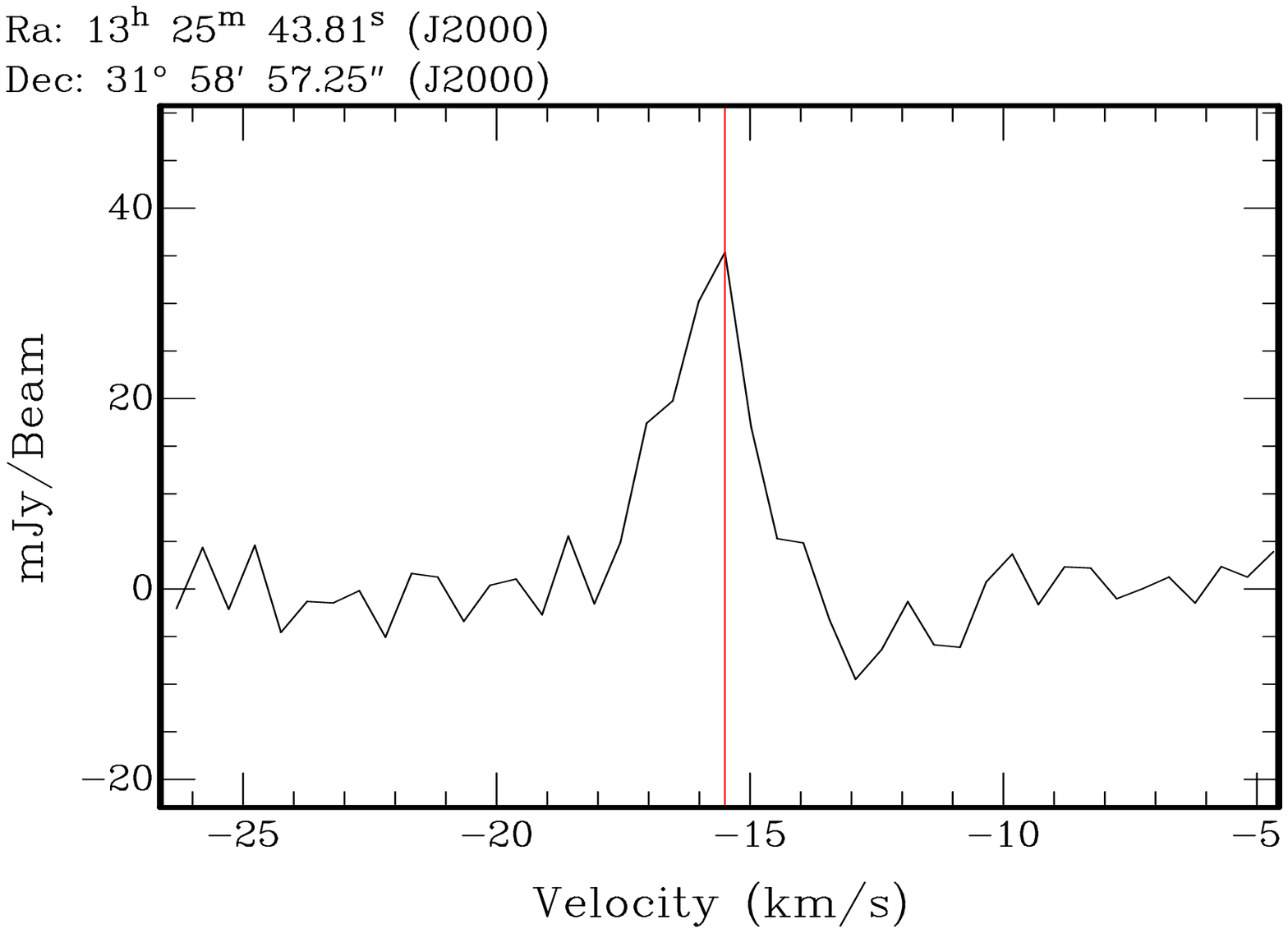}
\caption{(a) Top panel: \ion{H}{i} emission clump adjacent to 3C286 in map
  (left) and spectrum (right). (b) Bottom panel: \ion{H}{i} emission shell
  toward 4C+32.44 in map (left) and spectrum (right). 
}
\label{fig:3clump}
\end{figure}

The same WSRT observations that provided the absorption spectra also allow an
interferometric imaging search for emission counterparts.  Cubes of \ion{H}{i}
line emission were produced at a range of angular resolutions (15, 30, 60 and
120\arcsec). Obtaining sufficient brightness sensitivity to achieve detections
in emission typically required smoothing to 60\arcsec.  At those velocities
where total power emission exceeding about 1~K is seen in
Fig.~\ref{fig:allspec}, compact emission clumps of 2--3 K brightness are
detected at apparently random locations in the field, superposed on the
poorly-sampled (by the interferometer) diffuse background
emission. One such emission clump, immediately adjacent to the 3C286
line-of-sight is shown in Fig.~\ref{fig:3clump}(a). The intrinsic angular size
of these clumps appears to be about 30\arcsec, while their FWHM line-widths
are 1--2~km/s, corresponding to temperatures of $\sim$ 20--80~K. A
representative peak column density for the clumps in the 3C286 field is
N$_{\ion{H}{i}}~\sim~5 \times10^{18}$~cm$^{-2}$. At an assumed distance of,
say, 100~pc, the clumps would have a size of $\sim 3000$~AU, with a central
volume density of $\sim 100$~cm$^{-3}$.

An even more interesting emission structure is detected in the 4C+32.44
field. The line-of-sight toward this background source appears to intersect a
15\arcmin diameter shell of \ion{H}{i} emission, as shown in
Fig.\ref{fig:3clump}(b). Image quality is limited in
Fig.\ref{fig:3clump}(b) by a non-ideal configuration and no primary beam
correction has been applied. Although it
may be a chance superposition, this apparent shell contains the G0III star
HD~116856 at $(\alpha,\delta)_{2000}$ = (13:25:55.835,+31:51:40.629).  The
measured parallax of this star places it at $105 \pm 11$~pc, where the shell
would have a diameter of $\sim 0.45$~pc.  The stellar proper motion
$(\Delta\alpha,\Delta\delta)$ = (+14.73,$-$43.40)~mas/yr is directed
to the SE.  Peak N$_{\ion{H}{i}}~\sim~10^{19}$~cm$^{-2}$ is seen in this
structure, with FWHM line-widths of
2--3~km/s. It seems possible that this structure is associated with the
termination shock of a stellar wind.

\section{Cloud models}
\label{sec:models}

Various approaches are used in the literature to constrain physical conditions
from an analysis of \ion{H}{i} emission and absorption spectra (eg. Kanekar et
al.~\cite{kane03}, Heiles \& Troland~\cite{heil03}). An exhaustive
comparison of the various methods is beyond the scope of the current paper,
but will be addressed in Braun \& Kanekar~(\cite{brau05}). We simply note from
the outset that while it is always possible to model both \ion{H}{i} 
absorption and emission profiles as arising from the sum of multiple
Gaussians, it is far from clear whether a plausible combination of physical
conditions might exist along any real line-of-sight that would produce such an
artificially simplistic observable (ie. very long, yet iso-thermal,
and non-turbulent path-lengths). In fact, the most sensitive observations
of both emission and absorption lines (cf. Fig.~\ref{fig:allspec}) show
profiles that appear to be semi-Lorentzian, with narrow line-cores that merge
smoothly into broader wings. As more sensitive data become available, an 
increasing number of broader Gaussians becomes necessary to fit such profiles.

Motivated by the non-Gaussian line profiles we have explored some simple, 
spherically symmetric, isobaric two-phase cloud models of the form:
\begin{equation}
n_{\rm H}(r) = n_o {\rm exp}[-(r/s)^{\alpha_1}] \hskip1cm {\rm for~T~<~4000~K~and} 
\end{equation}
\begin{equation}
n_{\rm H}(r) = n_o {\rm exp}[-(r/s)^{\alpha_2}] \hskip1cm {\rm for~T~>~4000~K~~~~\ }
\end{equation}
where we further relate volume density to temperature using a constant thermal
pressure $P/k_B~=~n_{\rm H}T$ = 1500~cm$^{-3}$K.  The gas temperature was allowed to
vary between T$_{min} = 20~K$ and T$_{max} = 15000~K$, yielding an assumed
thermal velocity dispersion of $\sigma^2~=~0.0086 T$. The predicted \ion{H}{i}
absorption and total power emission spectra were calculated for a ``cloud''
placed at the central velocity of each observed feature in an attempt to
simultaneously reproduce the observed spectra shown in
Fig.~\ref{fig:allspec}. The most important free parameters in this process
were the cloud scale-length, $s$, the cloud distance, $d$ (which most strongly
influences the predicted total power emission) and an impact parameter,
$b$. This last parameter was used to allow for the likely circumstance that
each spherical model cloud may not be penetrated exactly on-axis by the
background absorber.  The two power-law indices of the scaled radius,
$\alpha_1$ and $\alpha_2$ determine the characteristic line shapes in the cold
core and warm halo respectively.  Although these were, in principle, also free
parameters, it was found that only minor variations from ``standard'' values
of about $\alpha_1$=1/4 and $\alpha_2$=1/8 were needed. Our arbitrary choice
of a transition temperature of 4000~K to separate the cool and warm phases is
also non-critical. Comparable fits are possible with other choices of the
transition temperature within a broad range. The broad-band ($\sim$ 90 km/s
FWHM) stray radiation contribution to the emission spectra is approximated by
a simple Gaussian, rather than a physical component.

\begin{figure}
\centering
\includegraphics[width=8.cm]{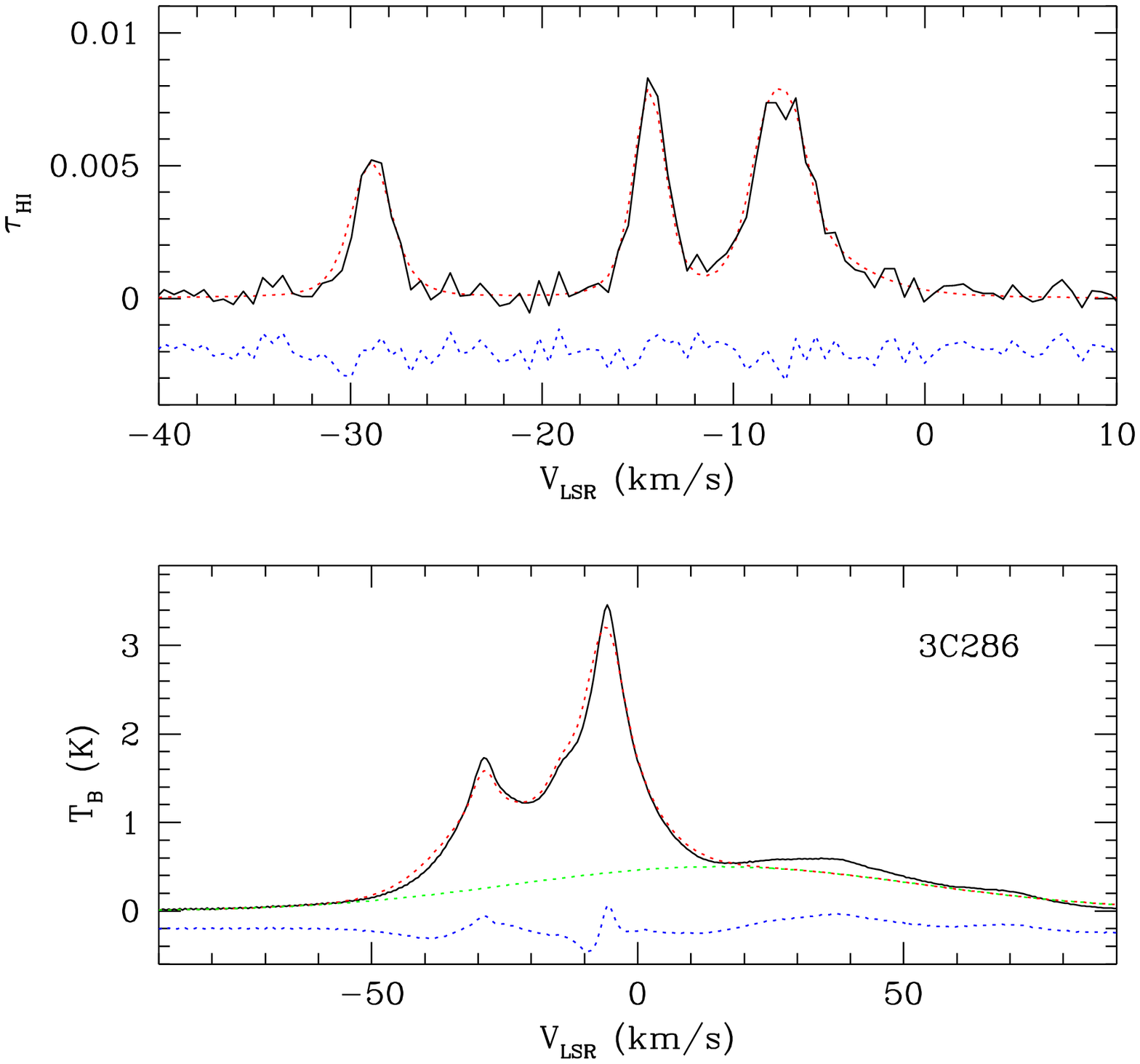}\\
\includegraphics[width=8.cm]{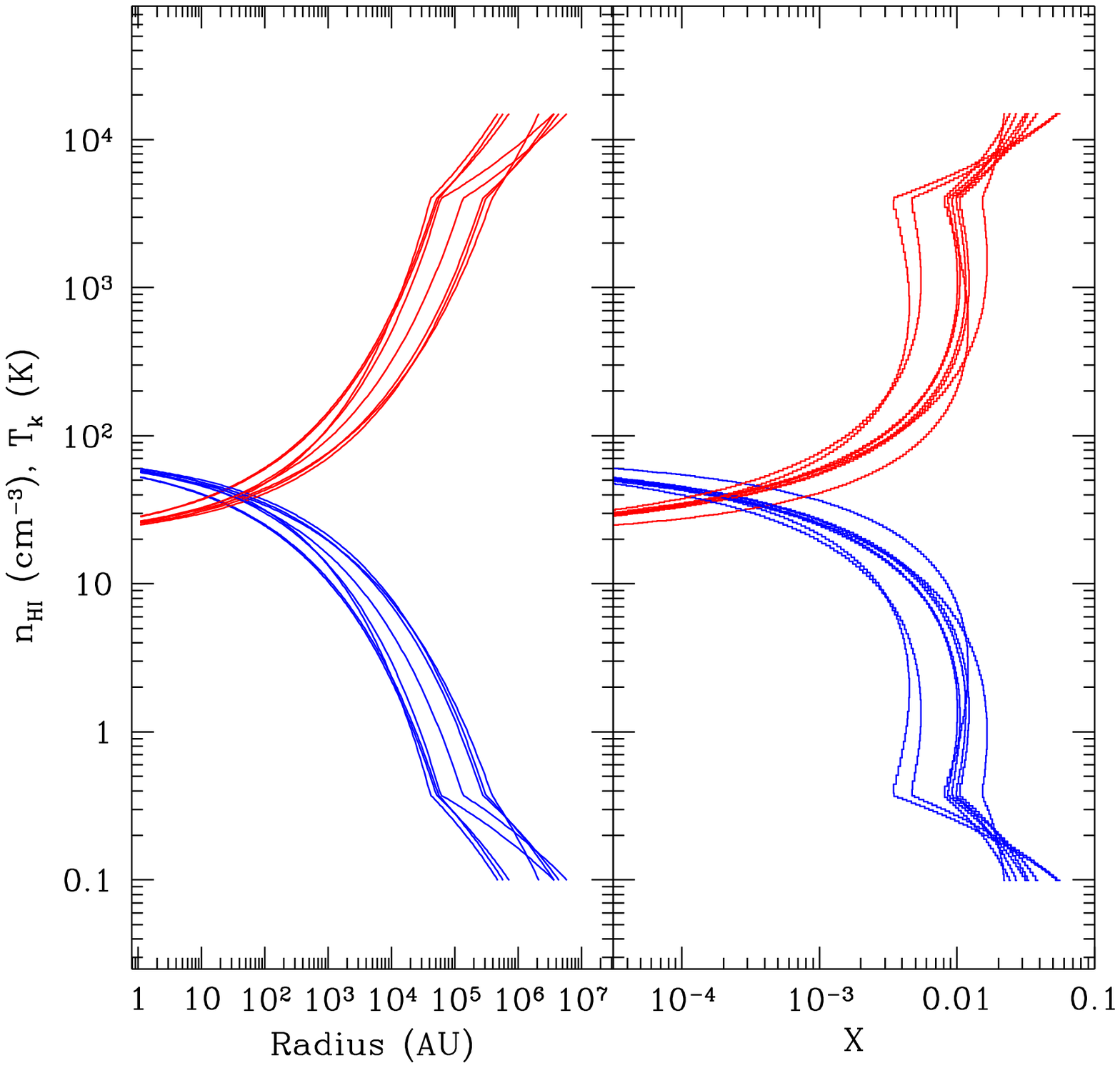}
\caption{Overlay of observed, modelled and residual spectra for the 3C286
  l-o-s (top). (The broad stray radiation contribution to the emission profile
  is plotted separately.) Density and temperature profiles (left) and
  fractional 
  distribution functions (right) of all modelled ``clouds'' (bottom).}
\label{fig:modcomp}
\end{figure}

An illustration of the simultaneous emission and absorption line fitting is
shown in the top panel of Fig.~\ref{fig:modcomp} for the 3C286 line-of-sight.
A formal least squares fit has not been carried out here, but merely a
$\chi$-by-eye to illustrate the possibilities of this approach. The bottom
panel of the figure shows the entire set of density and temperature profiles
as well as the fractional distribution functions for the modelled clouds along
all four lines-of-sight. This illustrates the basic similarities of the
modelled features, although the apparent scale-lengths do vary by about an
order of magnitude. The apparent distance used in reproducing the spectra was
about 10~pc, and the typical half-density radius was 100~AU, although these
were not very well-constrained given the very low angular resolution of our
total power data (35~arcmin). These modelling results can be easily scaled to
other assumed thermal pressures by a linear scaling of $n_{\rm H}$ together with an
inverse linear scaling of both $s$ and $d$. For example, with a typical
thermal pressure of only 150~cm$^{-3}$K, comparable fits would be obtained at
apparent distances of 100~pc and typical half-density radii of
1000~AU. Variations from the idealised spherical cloud symmetry would have a
similar impact on apparent distances and sizes. The preserved quantity in
these scalings is the distribution of column density over
temperature (or more precisely line-width). The fractional column density at a
particular density and temperature with respect to the total line-of-sight
column density is plotted in the lower right panel of
Fig.~\ref{fig:modcomp}. Column densities of the cool cores are typically only
a few percent of those in the warm envelopes. The total column densities of
the modelled clouds vary between 5--35$ \times10^{18}$~cm$^{-2}$.  While it
should be emphasised that the current attempts are fairly simplistic, it is
interesting that we obtain scale lengths quite similar to those of the
structures seen directly in emission in Fig~\ref{fig:3clump}. We also stress
that our cloud model ``temperatures'' may not be wholly thermal in nature, but
are likely to include a non-thermal, turbulent contribution.

\section{Discussion}

The narrow absorption features of Fig.~\ref{fig:allspec} have the lowest
column densities that have ever been detected in the CNM, two orders of
magnitude lower than the mean CNM column density $N_{\ion{H}{i}} = 2.7 \times
10^{20}$~cm$^{-2}$ of the ``classical'' McKee-Ostriker~(\cite{mcke77a}) model
(see also Stanimirovic \& Heiles \cite{stan05}). Our current results thus
suggest that even the most diffuse regions of the Galaxy are populated by
hitherto undiscovered, tiny distinct structures of very high density and
temperature contrast. This implies substantial injection of
fluctuation power on very small scales, whose source is presently unclear. The
detection of a narrow shell-like structure with small transverse separation
from a G0III star in one of our fields suggests that the stellar winds of
intermediate mass stars might play a role in the formation of these tiny
clumps, whenever such stars find themselves within a diffuse atomic
structure. More distributed sources of energy injection might also lead to
thermal condensation in a turbulent flow as discussed by Audit \& Hennebelle
(\cite{audi05}). These authors also demonstrate that local pressure
equilibrium is still approximately preserved in the vicinity of condensations,
despite a wide range of associated gas temperatures, consistent with the
isobaric two-phase cloud models we develop in \S\ref{sec:models}. The density
and temperature distribution functions of our model clouds,
Fig.~\ref{fig:modcomp} bottom right, also compare very favourably with those
of Audit \& Hennebelle (their Fig.8) in the highly turbulent regime.

We emphasise that the physical origin of these tiny \ion{H}{i}
structures and their general importance in the ISM is still unknown. Their
very low column densities should cause them to evaporate rapidly (e.g. McKee
\& Cowie~\cite{mcke77b}). Perhaps the first issue that needs to be clarified
is whether they are truly ubiquitous in the ISM. We plan to address this with
high sensitivity WSRT absorption spectra towards a larger sample of
bright high-latitude sources. Sensitive total power emission spectra with both
higher angular resolution and a lower stray-radiation contribution using the
Green Bank Telescope or Arecibo would also be invaluable. Clearly, more
work needs to be done to fully characterise the nature of sub-structure in the
diffuse ISM.

\begin{acknowledgements}
The WSRT is operated by ASTRON with support from the Netherlands Foundation
for Scientific Research (NWO).
\end{acknowledgements}

\end{document}